\documentclass [a4paper,fleqn, 12pt]{article}
\usepackage{graphicx}
\usepackage[small]{subfigure,epsfig}

\usepackage {amsmath} \usepackage{amssymb} \usepackage{cite}
\newcommand{\cn}

\begin{document}

%\begin{frontmatter}

\title
{Point vortices and classical orthogonal polynomials}

\author
{Maria V. Demina and Nikolay A. Kudryashov}

\date{Department of Applied Mathematics, National Research Nuclear University
MEPHI, 31 Kashirskoe Shosse,
115409 Moscow, Russian Federation}

%\author

%\corauthref{cor}},
%\corauth[cor]{Corresponding author.} \ead{nakudr@gmail.com}

%\address

\maketitle

\begin{abstract}

Stationary equilibria of point vortices with arbitrary choice of circulations in a background flow are studied. Differential equations satisfied by generating polynomials of vortex configurations are derived. It is shown that these equations can be reduced to a single one.  It is found that  polynomials that are Wronskians of classical orthogonal polynomials solve the latter equation. As a consequence vortex equilibria  at a certain choice of background flows can be described with the help of Wronskians of classical orthogonal polynomials.

\end{abstract}

%\begin{keyword}

% Point vortices, Special polynomials, Adler -- Moser polynomials

%\PACS 02.30.Hq - Ordinary differential equations

%\end{keyword}

%\end{frontmatter}

\section{Introduction}

The model of point vortices describing motion of two--dimensional incompressible fluid is one of the most elegant models of fluid dynamics. The motion of $M$ point vortices with circulations (or strengths)
$\Gamma_1$, $\ldots$~, $\Gamma_M$ at positions $z_1$, $\ldots$~, $z_M$ in zero background flow is governed by the Helmholtz's equations
\begin{equation*}
\label{Motion_of_Vortices}\frac{d z_k^{*}}{d\,t}=\frac{1}{2\pi i}\sum_{j=1}^{M}{}^{'}\frac{\Gamma_j}{z_k-z_j},\quad k=1,\ldots, M,
\end{equation*}
where the prime means that  the case $j=k$ is excluded and the symbol~$\,^{*}$ stands for complex conjugation. This system induces complicated dynamics and is not integrable in the case $M>3$. Nevertheless some particular "motions" including stationary, translating, and rotating equilibria are successfully studied \cite{Kadtke01, Aref01, Aref02, Aref03, Aref04, Aref05, Oneil01, Oneil02, Oneil03, Oneil04, Clarkson01, Borisov01, Demina25, Demina26}.  A convenient approach applicable to such types of motion is the so--called "polynomial method" \cite{Aref01}. According to this method polynomials with roots at vortex positions are introduced. This approach provides quite unexpected connection between dynamics of point vortices and the theory of classical and nonlinear special polynomials \cite{Aref01, Clarkson01, Demina25, Bartman01, Adler01, Demina15, Kudr08a}. For example, the generating polynomial of $M$ identical point vortices on a line in rotating relative equilibrium is essentially the $M$th Hermite polynomial. In $1964$ Tkachenko obtained a differential equation satisfied by generating polynomials of stationary vortex arrangements with equal in absolute value circulations \cite{Tkachenko01}. Now this equation is known as the Tkacheno equation. Later a generalization of the Tkachenko equation to the uniformly translating case was derived (for details see \cite{Kadtke01, Aref01, Aref05}). Again the latter equation describes translating equilibria of point vortices with equal in absolute value circulations. In recent work \cite{Demina26} it is shown that equations for generating polynomials of vortex arrangements with arbitrary choice of circulations can be reduced to the Tkachenko equation in the completely stationary case  and to the generalization of the the Tkachenko equation equation in the uniformly translating case. It turns out that the Adler -- Moser polynomials \cite{Adler01}, a famous sequence of polynomials, which generates polynomial solutions to the Tkachenko equation, provide not unique polynomial solutions of the latter equation. Some examples of alternative polynomial solutions are given in \cite{Demina26}.

In this article we study stationary equilibria of multivortex systems in a background flow. We derive differential equations satisfied by generating polynomials of vortices and reduce these equation to a simple form. We apply an approach suggested in \cite{Demina26}. Our aim is to show that vortex equilibria at a certain choice of background flows can be described with the help of polynomials that are Wronskians of classical orthogonal polynomials. These results provide additional link between the vortex theory and the the theory of classical orthogonal polynomials. We use the technic of Darboux transformations, for more details see \cite{Matveev01, Adler01, Oblomkov01}.

This article is organized as follows. In section \ref{Darboux_Orthogonal_Polynomials} we construct sequences of Darboux transformations for a second order linear differential equation and study in details the cases corresponding to the families of classical  orthogonal polynomials. In section  \ref{Background_flow} we consider stationary equilibrium of point vortices in a background flow and derive differential equations satisfied by generating polynomials of vortex configurations. We give background flows in explicit form for such cases that involve Wronskians of classical orthogonal polynomials. In addition in section \ref{Background_flow} we construct reductions of the differential equations satisfied by generating polynomials to a simple one. In section \ref{Orth_pols_examles} we study polynomial solutions of the latter equation and present several explicit examples.

\section{Darboux transformations for classical orthogonal polynomials} \label{Darboux_Orthogonal_Polynomials}

The systems of classical orthogonal polynomials can be constructed as polynomial solutions of the following second  order linear differential equation
\begin{equation}
\label{Ort_Pols_Equation}\sigma(z)\psi_{zz}+\tau(z)\psi_z+\lambda \psi=0.
\end{equation}
In this expression $\sigma(z)$ is a polynomial of the degree at most two, $\tau(z)$ is a polynomial of the degree at most one, and $\lambda$ is a real constant. First of all we shall  consider a more general second  order equation
\begin{equation}
\label{Ort_Pols_Equation_Gen}\sigma(z)\psi_{zz}+\tau(z)\psi_z+\{u(z)+\lambda\} \psi=0
\end{equation}
with $\sigma(z)$, $\tau(z)$, $u(z)$ being sufficiently smooth functions. Let us denote a nontrivial solution of equation \eqref{Ort_Pols_Equation_Gen} with the parameter $\lambda=\lambda_1$ as $\psi_1$.

{\textbf{Theorem 2.1.}}
\emph{For any solution $\psi$ of equation \eqref{Ort_Pols_Equation_Gen} the Darboux transformation
\begin{equation}
\label{DT1}\tilde{\psi}=\psi_z-\frac{\psi_{1,z}}{\psi_1}\psi,
\end{equation}
gives a solution of the following equation
\begin{equation}
\label{Ort_Pols_Equation_Gen_DT1}\sigma(z)\tilde{\psi}_{zz}+\tau_1(z)\tilde{\psi}_z+\{u_1(z)+\lambda\} \tilde{\psi}=0,
\end{equation}
where the functions $\tau_1(z)$, $u_1(z)$ are given by
\begin{equation}\begin{gathered}
\label{Tau_U_1}\tau_1(z)=\tau(z)+\sigma_z(z),\quad u_1(z)=u(z)+\tau_z(z)+\sigma_z(z)\{\ln \psi_1\}_z+\\
+2\sigma(z)\{\ln \psi_1\}_{zz}.
\end{gathered}\end{equation}}

\textbf{Proof.}
We introduce the second order operators $L$ and $L_1$ according to the rules
\begin{equation}
\label{Operators_L}L\stackrel{def}{=}\sigma(z)D^2+\tau(z)D+u(z),\quad L_1\stackrel{def}{=}\sigma(z)D^2+\tau_1(z)D+u_1(z),\quad D\stackrel{def}{=}\frac{d}{d z}
\end{equation}
and rewrite equations \eqref{Ort_Pols_Equation_Gen}, \eqref{Ort_Pols_Equation_Gen_DT1} in the following way
\begin{equation}
\label{Eqs_L}(L+\lambda)\psi=0,\quad (L_1+\lambda)\tilde{\psi}=0.
\end{equation}
Substituting the transformation \eqref{DT1}, which we rewrite in the form
\begin{equation}
\label{DT1_A}\tilde{\psi}=A_{\psi_1}\psi,\quad A_{\psi_1}\stackrel{def}{=} D -\frac{\psi_{1,z}}{\psi_1},
\end{equation}
into the second equation in \eqref{Eqs_L} and using the first equation in \eqref{Eqs_L}, we obtain the relation
\begin{equation}
\label{DT_Relation}(L_1 A_{\psi_1}-A_{\psi_1}L_1)\psi=0.
\end{equation}
Expression \eqref{DT_Relation} is a first order polynomial in $\psi_{zz}$, $\psi_{z}$, $\psi$. Setting to zero coefficients at $\psi_{zz}$, $\psi_{z}$ yield relations \eqref{Tau_U_1} for the functions $\tau_1(z)$, $u_1(z)$  accordingly. Using these relations and the equation for the function $\psi_1$, we see that the coefficient at $\psi$ in expression \eqref{DT_Relation} vanishes. This completes the proof.

Further we note that the Darboux transformation \eqref{DT1} can be iterated. As a result we obtain the following theorem.

\textbf{Theorem 2.2.}
\emph{Let $\psi_1$, $\ldots$, $\psi_k$ be nontrivial solutions of equation \eqref{Ort_Pols_Equation_Gen} with pairwise different values of the parameter $\lambda$: $\lambda_1$, $\ldots$, $\lambda_k$. Then the Darboux transformation
\begin{equation}
\label{DTk}\tilde{\psi}=\frac{W[\psi_1, \ldots, \psi_k,\psi]}{W[\psi_1, \ldots, \psi_k]}
\end{equation}
gives a solution of the equation
\begin{equation}
\label{Ort_Pols_Equation_Gen_DTk}\sigma(z)\tilde{\psi}_{zz}+\tau_k(z)\tilde{\psi}_z+\{u_k(z)+\lambda\} \tilde{\psi}=0
\end{equation}
whenever the function $\psi$ solves equation \eqref{Ort_Pols_Equation_Gen}. In this expressions $W\stackrel{def}{=}W[\psi_1, \ldots, \psi_k]$ is the Wronskian of the functions $\psi_1$, $\ldots$, $\psi_k$ and
\begin{equation}
\begin{gathered}
\label{Tau_U_k}\tau_k(z)=\tau(z)+k\sigma_z(z),\hfill \\
u_k(z)=u(z)+k\tau_z(z)+\frac{k(k-1)}{2}\sigma_{zz}(z)+\sigma_z(z)\{\ln W\}_z+\\
+2\sigma(z)\{\ln W\}_{zz}.
\end{gathered}
\end{equation}}

\textbf{Proof.}
Applying $k$ times the Darboux transformations to a solution $\psi$ of equation \eqref{Ort_Pols_Equation_Gen}, we see that the function
\begin{equation}
\label{Sol_DTk}\tilde{\psi}=M_k\psi,\quad M_k\stackrel{def}{=}A_{f_k}\ldots A_{f_{2}}A_{f_1},\quad f_1=\psi_1
\end{equation}
satisfies equation \eqref{Ort_Pols_Equation_Gen_DTk} with $\tau_k(z)$, $u_k(z)$ given by
\begin{equation}
\begin{gathered}
\label{Tau_U_k_Proof}\tau_k(z)=\tau(z)+k\sigma_z(z),\hfill \\
u_k(z)=u(z)+k\tau_z(z)+\frac{k(k-1)}{2}\sigma_{zz}(z)+\sigma_z(z)\{\ln f_1\ldots f_k\}_z+\\
+2\sigma(z)\{\ln f_1\ldots f_k\}_{zz}.
\end{gathered}
\end{equation}
We note that at the $n^{\text{th}}$ step, $n=1$, $\ldots$, $k$ we use the Darboux transformation build on the basis of the function $f_n$. Thus we obtain the set of the functions $f_1$, $\ldots$, $f_k$
\begin{equation}\begin{gathered}
\label{f_DTk}f_1=\psi_1,\, f_2=A_{f_1}\psi_2,\, \ldots,f_n=A_{f_{n-1}}\ldots A_{f_{2}}A_{f_1}\psi_n,\, \ldots, \, f_k=\\
=A_{f_{k-1}}\ldots A_{f_{2}}A_{f_1}\psi_k
\end{gathered}\end{equation}
From our construction it follows that the functions $\psi_1$, $\ldots$, $\psi_n$ provide $n$ linearly independent solutions of the $n^{\text{th}}$ order linear differential equation
\begin{equation}
\label{M_DTk}M_n w=0,\quad M_n\stackrel{def}{=}A_{f_n}\ldots A_{f_{2}}A_{f_1}.
\end{equation}
Further we rewrite this equation in explicit form
\begin{equation}
\label{f_DTk_Coeff}D^nw+m_{n-1}D^{n-1}w+\ldots +m_0 w=0,\quad M_n=D^n+m_{n-1}D^{n-1}+\ldots +m_0.
\end{equation}
Substituting the functions $\psi_1$, $\ldots$, $\psi_n$ into equation \eqref{f_DTk_Coeff}, we get the system of linear algebraic equations with respect to the coefficients $m_0$, $\ldots$, $m_{n-1}$
\begin{equation}
\label{Algbraic_System_DTk}m_0\psi_l+m_1 D\psi_l+\ldots +m_{n-1}D^{n-1}\psi_l=-D^n\psi_l\quad l=1,\ldots , n.
\end{equation}
 Solving this system with the help of the Kramer's rule yields
\begin{equation}
\label{Algbraic_System_Solution}m_{l-1}=-\frac{W_l}{W[\psi_1,\ldots,\psi_n]}\quad l=1,\ldots , n,
\end{equation}
where the determinant $W_l$ is obtained by replacing the $l^{\text{th}}$ column of the Wronskian $W[\psi_1,\ldots$~, $\psi_n]$ by the column of the functions $D^n \psi_1$, $\ldots$, $D^n \psi_n$. Consequently, we see that the operator $M_n$ acts on any function $w$ as
\begin{equation}
\label{M_n_w}M_n w=\frac{W[\psi_1,\ldots ,\psi_n,w]}{W[\psi_1,\ldots ,\psi_n]}
\end{equation}
Now we return to Darboux transformation \eqref{Sol_DTk}. It follows from relation \eqref{M_n_w} that transformation \eqref{Sol_DTk} is given by \eqref{DTk}. Expressions \eqref{Tau_U_k_Proof} are obtained by induction applying $k$ times formulae \eqref{Tau_U_1}. It remains to find the product $f_1$ $\ldots$ $f_k$. Using expressions \eqref{f_DTk}, \eqref{M_DTk}, \eqref{M_n_w}, we obtain
\begin{equation}
\label{Product_f}f_1 \ldots f_k=\psi_1\, \frac{W[\psi_1,\psi_2]}{\psi_1}\, \ldots \frac{W[\psi_1,\ldots,\psi_k]}{W[\psi_1,\ldots,\psi_{k-1}]}=W[\psi_1,\ldots,\psi_k].
\end{equation}
Substituting relation \eqref{Product_f} into expressions \eqref{Tau_U_k_Proof}, we see that the function \eqref{DTk} is a solution of equation \eqref{Ort_Pols_Equation_Gen_DTk} provided that conditions \eqref{Tau_U_k} hold.

Now let us apply theorem 2.2 to the equation for classical orthogonal polynomials (see \eqref{Ort_Pols_Equation}). Suppose $p_{i_1}$, $\ldots$, $p_{i_k}$, $p_{i_{k+1}}$ are classical orthogonal polynomials satisfying equation \eqref{Ort_Pols_Equation} with pairwise different values of the parameter $\lambda$: $\lambda_{i_1}$ $\ldots$, $\lambda_{i_k}$, $\lambda_{i_{k+1}}$. Then the polynomials
\begin{equation}
\label{Determinants_Orthogonal_Polynomials}P(z)=W[p_{i_1},\ldots,p_{i_k},p_{i_{k+1}}],\quad Q(z)=W[p_{i_1},\ldots,p_{i_k}]
\end{equation}
solve the bilinear equation
\begin{equation}
\begin{gathered}
\label{Eqn_Determinants_Orthogonal_Polynomials}\sigma\left\{P_{zz}Q-2P_zQ_z+PQ_{zz}\right\}+\left[\tau+\left(k-\frac12\right)\sigma_z\right]\left\{P_zQ-PQ_z\right\}
+ \\
+\frac{\sigma_z}{2}\{P_zQ+PQ_z\}
+\left[\frac{k(k-1)}{2}\sigma_{zz}+k\tau_z+\lambda\right]PQ=0,\quad \lambda=\lambda_{i_{k+1}}.
\end{gathered}
\end{equation}
This result is obtained substituting $\tilde{\psi}=P/Q$, $u(z)=0$, and expressions \eqref{Tau_U_k} with $W=Q(z)$ into equation \eqref{Ort_Pols_Equation_Gen_DTk}, where we set $\lambda=\lambda_{i_{k+1}}$. Further finding the highest powers of Wronskians, we obtain the degrees of the polynomials $P(z)$, $Q(z)$
\begin{equation}
\label{Degreees_W_Orthogonal_Polynomials}\deg P=\sum_{l=1}^{k+1}i_l-\frac{k(k+1)}{2},\quad \deg Q=\sum_{l=1}^{k}i_l-\frac{k(k-1)}{2},
\end{equation}
where $i_l$ is the degree of the polynomial $p_{i_l}$. In the next section our aim is to show that equation \eqref{Eqn_Determinants_Orthogonal_Polynomials} can be obtained in the framework of the vortex theory.

\section{Point vortices in a background flow} \label{Background_flow}

The motion of a multivortex system in a background flow $w(z)$ is described by the  Helmholtz's equations with additional term
\begin{equation}
\label{Motion_of_Vortices_Background_Flow}\frac{d z_k^{*}}{d\,t}=\frac{1}{2\pi i}\sum_{j=1}^{M}{}^{'}\frac{\Gamma_j}{z_k-z_j}+\frac{w^{*}(z_k)}{2\pi i},\quad k=1,\ldots, M.
\end{equation}
In what follows we take the background flow in the form
 \begin{equation}
\label{Background_Flow_Expression}w^{*}(z_k)=\frac{U(z_k)}{2R(z_k)}+\frac{\Gamma_k}{4}\frac{ R_z(z_k)}{R(z_k)},
\end{equation}
where $R(z)$ is a polynomial of degree at most two and $U(z)$ is a polynomial of degree at most one. We exclude the case when $R(z_{k_0})=0$ unless $z=z_{k_0}$ is a simple root of the polynomial $R(z)$ and $U(z_{k_0})+\Gamma_{k_0}R_z(z_{k_0})/2=0$. Let us subdivide the vortices into groups according to the values of their circulations $\Gamma_j$, $j=1$,~$\ldots$~,~$N$. In other words, we suppose that there are $N$ different values of circulations in the arrangements we study. By $a_{1}^{(j)}$,~$\ldots$~,~$a_{l_j}^{(j)}$ we denote the positions of the vortices with circulation
 $\Gamma_j$. Therefore, we have $M=l_1$ $+$ $\ldots$ $+$ $l_N$. Let us introduce the polynomials
 \begin{equation}
\label{Polynoials_at_positions_of_vortices_multi_BF}P_j(z)=\prod_{i=1}^{l_j}(z-a_{i}^{(j)}),\quad j=1, \ldots, N.
\end{equation}
with roots at the vortex positions. We shall consider the stationary case, this yields $d z_k^{*}/ dt =0$. From equations \eqref{Motion_of_Vortices_Background_Flow} we find the system of algebraic relations
\begin{equation}
\begin{gathered}
\label{Relations_for_positions_of_vortices_Background_Flow}\sum_{j=1}^{N}\sum_{i=1}^{l_j}{}^{'}\frac{\Gamma_j}{a-a_{i}^{(j)}} +\frac{U\left(a\right)}{2R\left(a\right)}+\frac{\Gamma_{j_0}}{4}\frac{ R_z\left(a\right)}{R\left(a\right)}=0,\,\\
a=a_{i_0}^{(j_0)},\, i_0=1,\ldots, l_{j_0}, \, j_0=1,\ldots, N,
\end{gathered}
\end{equation}
where the case $(j_0,i_0)=(j,i)$ in summation is excluded. Using properties of the logarithmic derivative, we obtain the following equalities
\begin{equation}
\label{Polynoials_at_positions_of_vortices_Derivatives}P_{j,z}=P_j\sum_{i=1}^{l_j}\frac{1}{z-a_{i}^{(j)}},\quad P_{j,zz}=2P_j\sum_{i=1}^{l_j}\sum_{k=1}^{l_j}{}^{'}\frac{1}{(z-a_{i}^{(j)})(a_{i}^{(j)}-a_{k}^{(j)})}.
\end{equation}
Now let $z$ tend to one of the roots $a=a_{i_0}^{(j_0)}$ of the polynomial $P_{j_0}(z)$. Calculating the limit $z \rightarrow a$ in the expression for $P_{j_0,zz}$ yields
\begin{equation}
\label{Derivatives_a_j}P_{j_0,zz}(a)=2P_{j_0,z}(a)\sum_{i=1}^{l_{j_0}}{}^{'}\frac{1}{a-a_{i}^{(j_0)}}.
\end{equation}
Using equalities \eqref{Relations_for_positions_of_vortices_Background_Flow}, \eqref{Polynoials_at_positions_of_vortices_Derivatives}, we get the conditions
\begin{equation}
\label{Conditions_at_roots}\Gamma_{j_0}\frac{P_{j_0,zz}(a)}{P_{j_0,z}(a)}=-2\sum_{j=1,\,j\neq j_0}^{N}\Gamma_j \frac{P_{j,z}(a)}{P_{j}(a)} -\frac{U\left(a\right)}{R\left(a\right)}-\frac{\Gamma_{j_0}}{2}\frac{ R_z\left(a\right)}{R\left(a\right)},
\end{equation}
which are valid for any root $a=a_{i_0}^{(j_0)}$  of the polynomial $P_{j_0}(z)$. Further we see that the polynomial
\begin{equation}\begin{gathered}
\label{Polynomial_for_vort}\prod_{i=1}^{N}P_i(z)\left\{R(z)\sum_{j=1}^{N}\Gamma_j^2\frac{P_{j,zz}}{P_j}+2R(z)\sum_{k<j}\Gamma_k\Gamma_j\frac{P_{k,z}}{P_k}\frac{P_{j,z}}{P_j}
+\right.\\ \left. +\sum_{j=1}^{N}\left(\Gamma_jU(z)+\frac{\Gamma_j^2}{2}R_z(z)\right)\frac{P_{j,z}}{P_j} \right\}
\end{gathered}\end{equation}
 being of degree $M+\max\{\deg R, \deg U+1\}-2$ possesses the same roots as the polynomial $P_1(z)\times \ldots \times P_N(z)$. Thus we find the equation
\begin{equation}\begin{gathered}
\label{Equation_for_vort_Background_Flow}R(z)\sum_{j=1}^{N}\Gamma_j^2\frac{P_{j,zz}}{P_j}+2R(z)\sum_{k<j}\Gamma_k\Gamma_j\frac{P_{k,z}}{P_k}\frac{P_{j,z}}{P_j}
+\\
+\sum_{j=1}^{N}\left(\Gamma_jU(z)+\frac{\Gamma_j^2}{2}R_z(z)\right)\frac{P_{j,z}}{P_j} -C=0,
\end{gathered}\end{equation}
where the constant $C$ is obtained setting to zero the coefficient at $z^0$ in the Laurent expansion of the left--hand side in the relation \eqref{Equation_for_vort_Background_Flow}:
\begin{equation}
\label{Const_EQ_Background_Flow}C=\frac{R_{zz}(z)}{2}\left(\sum_{j=1}^{N}\Gamma_j\deg P_j\right)^2+U_z(z)\sum_{j=1}^{N}\Gamma_j\deg P_j.
\end{equation}
Further following an approach suggested in the article \cite{Demina26}, we introduce new functions $\tilde{P}(z)$, $\tilde{Q}(z)$ according to the rules
\begin{equation}
\begin{gathered}
\label{New_Pols_Arbitrary_Circ}\tilde{P}(z)=\prod_{j=1}^{N}P_j^{\frac{\Gamma_j(\Gamma_j+1)}{2}}(z),\quad \tilde{Q}(z)=\prod_{j=1}^{N}P_j^{\frac{\Gamma_j(\Gamma_j-1)}{2}}(z).
\end{gathered}
\end{equation}
Using the equalities
\begin{equation}
\begin{gathered}
\label{Relations_Tilde}\frac{d^2}{dz^2}\ln\left\{\tilde{P}(z)\tilde{Q}(z)\right\}=\sum_{j=1}^{N}\Gamma_j^2\left(\frac{P_{j,zz}}{P_j}-\frac{P_{j,z}^2}{P_j^2}\right), \\
\quad
\frac{d}{dz}\ln\left\{\frac{\tilde{P}(z)}{\tilde{Q}(z)}\right\}=\sum_{j=1}^{N}\Gamma_j\frac{P_{j,z}}{P_j},
\end{gathered}
\end{equation}
we get the differential equation for the functions $\tilde{P}(z)$ and $\tilde{Q}(z)$ of the form
\begin{equation}
\begin{gathered}
\label{Equation_Tilde_PQ_BF_Alt}R(z)\frac{d^2}{dz^2}\ln\left\{\tilde{P}(z)\tilde{Q}(z)\right\}+R(z)\left(\frac{d}{dz}\ln\left\{\frac{\tilde{P}(z)}{\tilde{Q}(z)}\right\}\right)^2 + U(z)\frac{d}{dz}\ln\left\{\frac{\tilde{P}(z)}{\tilde{Q}(z)}\right\}\\
 +\frac{R_z(z)}{2}\frac{d}{dz}\ln\left\{\tilde{P}(z)\tilde{Q}(z)\right\}-C=0.
\end{gathered}
\end{equation}
Note that this equation can be rewritten in the bilinear form as
\begin{equation}
\begin{gathered}
\label{Equation_Tilde_PQ_BF}R(z)\left\{\tilde{P}_{zz}\tilde{Q}-2\tilde{P}_z\tilde{Q}_z+
\tilde{P}\tilde{Q}_{zz}\right\}+U(z)\left\{\tilde{P}_z\tilde{Q}-\tilde{P}\tilde{Q}_z\right\}+ \\
+\frac{R_z(z)}{2}\left\{\tilde{P}_z\tilde{Q}+\tilde{P}\tilde{Q}_z\right\}-C\tilde{P}\tilde{Q}=0.
\end{gathered}
\end{equation}
If $\tilde{P}$, $\tilde{Q}$ in \eqref{Equation_Tilde_PQ_BF} are polynomials, then for the constant $C$ we obtain
\begin{equation}
\label{Const_EQ_Background_Flow_Tilde}C=\frac{R_{zz}(z)}{2}\left(\deg\tilde{P} - \deg\tilde{Q} \right)^2+U_z(z)\left(\deg\tilde{P} - \deg\tilde{Q}\right).
\end{equation}
Thus we see that stationary equilibria of point vortices with circulations $\Gamma_1$, $\ldots$, $\Gamma_N$ in the background flow given by \eqref{Background_Flow_Expression} can be described with the help of equation \eqref{Equation_Tilde_PQ_BF}. Suppose we have found a solution of equation \eqref{Equation_Tilde_PQ_BF} in the form \eqref{New_Pols_Arbitrary_Circ}, then  a vortex with circulation $\Gamma_j$ is situated at the point $z=z_0$ whenever  the function $\tilde{P}(z)$ has a "root" of "multiplicity" $\Gamma_j(\Gamma_j+1)/2$ at the point $z=z_0$ and the function $\tilde{Q}(z)$ has a "root" of "multiplicity" $\Gamma_j(\Gamma_j-1)/2$ at the point $z=z_0$. The circulation is calculated as the difference of the corresponding "multiplicities". We say that the function $f(z)$ has a "root" of "multiplicity" $k$ at the point $z=z_0$ if $f(z)=(z-z_0)^k \psi(z)$ with $\psi(z)$ being an analytic function (possibly multivalued)  in a neighborhood of $z_0$ and $\psi(z_0)\neq 0$. Suppose $k \in \mathbb{N}$, then the point $z=z_0$ is a root of the function $f(z)$ in the usual cense.

Now we shall apply our results to the following multivortex situation. Suppose $l_n^{+}$ vortices with circulations $n\Gamma$ are situated at positions $z=a_{i}^{(n)}$, $i=1$, $\ldots$, $l_n^{+}$, $n=1$, $\ldots$, $N_1$ and $l_m^{-}$ vortices with circulations $-m\Gamma$ are situated at positions $z=b_{k}^{(m)}$, $k=1$, $\ldots$, $l_m^{-}$, $m=1$, $\ldots$, $N_2$. Further we consider the polynomials
\begin{equation}\begin{gathered}
\label{Polynoials_at_positions_of_vortices_PQ}P_n(z)=\prod_{i=1}^{l_n^{+}}(z-a_{i}^{(n)}),\quad n=1,\ldots , N_1 \hfill \\
 Q_m(z)=\prod_{k=1}^{l_m^{-}}(z-b_{k}^{(m)}),\quad m=1,\ldots , N_2
\end{gathered}
\end{equation}
with no common and multiple roots. The amount of vortices in such an arrangement is equal to $M=l_1^{+}+$ $\ldots$ $+$ $l_{N_1}^{+}$ $+$ $l_1^{-}$ $+$ $\ldots$ $+$ $l_{N_2}^{-}$. We see that stationary equilibria of this multivortex system in the background flow  \eqref{Background_Flow_Expression} can be described in terms of the functions
\begin{equation}
\begin{gathered}
\label{New_Pols_Addition}\tilde{P}(z)=\prod_{n=1}^{N_1}P_n^{\frac{\Gamma n(\Gamma n+1)}{2}}(z)\prod_{m=1}^{N_2}Q_m^{\frac{\Gamma m(\Gamma m-1)}{2}}(z),\, \\ \tilde{Q}(z)=\prod_{n=1}^{N_1}P_n^{\frac{\Gamma n(\Gamma n-1)}{2}}(z)\prod_{m=1}^{N_2}Q_m^{\frac{\Gamma m(\Gamma m+1)}{2}}(z),
\end{gathered}
\end{equation}
which satisfy equation \eqref{Equation_Tilde_PQ_BF}. Along with this we can consider the polynomials of the form
\begin{equation}
\begin{gathered}
\label{New_Pols}\tilde{P}(z)=\prod_{n=1}^{N_1}P_n^{\frac{n(n+1)}{2}}(z)\prod_{m=1}^{N_2}Q_m^{\frac{m(m-1)}{2}}(z),\quad \tilde{Q}(z)=\prod_{n=1}^{N_1}P_n^{\frac{n(n-1)}{2}}(z)\prod_{m=1}^{N_2}Q_m^{\frac{m(m+1)}{2}}(z),
\end{gathered}
\end{equation}
and conclude that these polynomials satisfy the equation
\begin{equation}
\begin{gathered}
\label{Equation_Tilde_PQ_BF_ALT}R(z)\left\{\tilde{P}_{zz}\tilde{Q}-2\tilde{P}_z\tilde{Q}_z+
\tilde{P}\tilde{Q}_{zz}\right\}+\frac{U(z)}{\Gamma}\left\{\tilde{P}_z\tilde{Q}-
\tilde{P}\tilde{Q}_z\right\}+ \\
+\frac{R_z(z)}{2}\left\{\tilde{P}_z\tilde{Q}+\tilde{P}\tilde{Q}_z\right\}-C_1\tilde{P}\tilde{Q}=0.
\end{gathered}
\end{equation}
From expression \eqref{Const_EQ_Background_Flow_Tilde} it follows that
\begin{equation}
\label{Const_EQ_Background_Flow_Tilde_ALT}C_1=\frac{R_{zz}(z)}{2}\left(\deg\tilde{P} - \deg\tilde{Q} \right)^2+\frac{U_z(z)}{\Gamma}\left(\deg\tilde{P} - \deg\tilde{Q}\right).
\end{equation}

Now recalling the results of section \ref{Darboux_Orthogonal_Polynomials}, we obtain that equation \eqref{Equation_Tilde_PQ_BF} coincides with equation \eqref{Eqn_Determinants_Orthogonal_Polynomials} provided that
\begin{equation}
\begin{gathered}
\label{BF_exact}R(z)=\sigma(z),\quad  U(z)=\tau(z)+\left(k-\frac12\right)\sigma_z(z)
\end{gathered}
\end{equation}
and $\tilde{P}(z)=P(z)$, $\tilde{Q}(z)=Q(z)$. By direct calculations we verify that the coefficient at $\tilde{P} \tilde{Q}$ in \eqref{Equation_Tilde_PQ_BF} coincides with the coefficient at $PQ$ in \eqref{Eqn_Determinants_Orthogonal_Polynomials} whenever relation \eqref{BF_exact} is valid. Consequently, the polynomials
\begin{equation}
\label{Equation_Tilde_PQ_BF_ES}\tilde{P}(z)=W[p_{i_1},\ldots,p_{i_k},p_{i_{k+1}}],\quad \tilde{Q}(z)=W[p_{i_1},\ldots,p_{i_k}]
\end{equation}
where $p_{i_1}$, $\ldots$ , $p_{i_{k+1}}$ are pairwise different classical orthogonal polynomials, solve equation \eqref{Equation_Tilde_PQ_BF} provided that relations \eqref{BF_exact} hold.

\section{Hermite, Laguerre, and Jacobi polynomials in the framework of the vortex theory} \label{Orth_pols_examles}

In this section we study polynomial solutions of equation \eqref{Equation_Tilde_PQ_BF} and give several explicit examples.
We begin with the theorem.

 \textbf{Theorem 4.1.}
\emph{ Suppose a pair of polynomials $\tilde{P}(z)$, $\tilde{Q}(z)$ satisfy equation \eqref{Equation_Tilde_PQ_BF} and $z=z_0$ is such a point in the complex plane that $R(z_0)\neq 0$; then the  following statements are valid:
 \begin{enumerate}
\item If the point $z_0$ is a multiple root of one of the polynomials $\tilde{P}(z)$, $\tilde{Q}(z)$, then it is also a root of another polynomial and the multiplicities of the root $z_0$ for the polynomials $\tilde{P}(z)$, $\tilde{Q}(z)$ are two successive triangular numbers.
\item If the point $z_0$ is a common root of the polynomials $\tilde{P}(z)$, $\tilde{Q}(z)$, then it is a multiple root of at least one of them  and the multiplicities of the root $z_0$ for the polynomials $\tilde{P}(z)$, $\tilde{Q}(z)$ are two successive triangular numbers.
\end{enumerate}}

\textbf{Proof.}
In order to prove statements 1, 2 of the theorem we substitute the expressions
\begin{equation}
\begin{gathered}
\label{BF_roots_T}\tilde{P}(z)=(z-z_0)^rh(z),\quad \tilde{Q}(z)=(z-z_0)^sg(z),\quad r,s\in \mathbb{N}\cup\{0\},
\end{gathered}
\end{equation}
where $h(z)$, $g(z)$ are polynomials such that $h(z_0)\neq 0$, $g(z_0)\neq 0$ into equation  \eqref{Equation_Tilde_PQ_BF} and find the Tailor series in a neighborhood of the point $z_0$ of the resulting relation. Setting to zero the coefficient at $(z-z_0)^{r+s-2}$, we obtain the algebraic equation
\begin{equation}
\begin{gathered}
\label{Multiplicities_Relation_T}(r-s)^2=r+s.
\end{gathered}
\end{equation}
Thus we get the system
\begin{equation}
\begin{gathered}\label{Multiplicities_Relation_System_T}r-s=l,\quad r+s=l^2,\quad l\in \mathbb{Z}.
\end{gathered}
\end{equation}
Solving this system yields
\begin{equation}
\begin{gathered}\label{Multiplicities_Res_T}r=\frac{l(l+1)}{2},\quad s=\frac{l(l-1)}{2},\quad l\in \mathbb{Z}.
\end{gathered}
\end{equation}
Analyzing expressions \eqref{Multiplicities_Res_T}, we make sure that statements 1, 2 are valid.

As a consequence of theorem 4.1 we see that any polynomial solution $\tilde{P}(z)$, $\tilde{Q}(z)$ of equation \eqref{Equation_Tilde_PQ_BF} such that the polynomials $\tilde{P}(z)$, $\tilde{Q}(z)$ do not have common roots with the polynomial $R(z)$ describes equilibrium of point vortices in a background flow. Indeed, from theorem 4.1 it follows that the polynomial solution in question  can be always presented in the form \eqref{New_Pols_Arbitrary_Circ}.

\begin{figure}[t]
 \centerline{
 \subfigure[$LR_{5}^{(2)}(z)$]{\epsfig{file=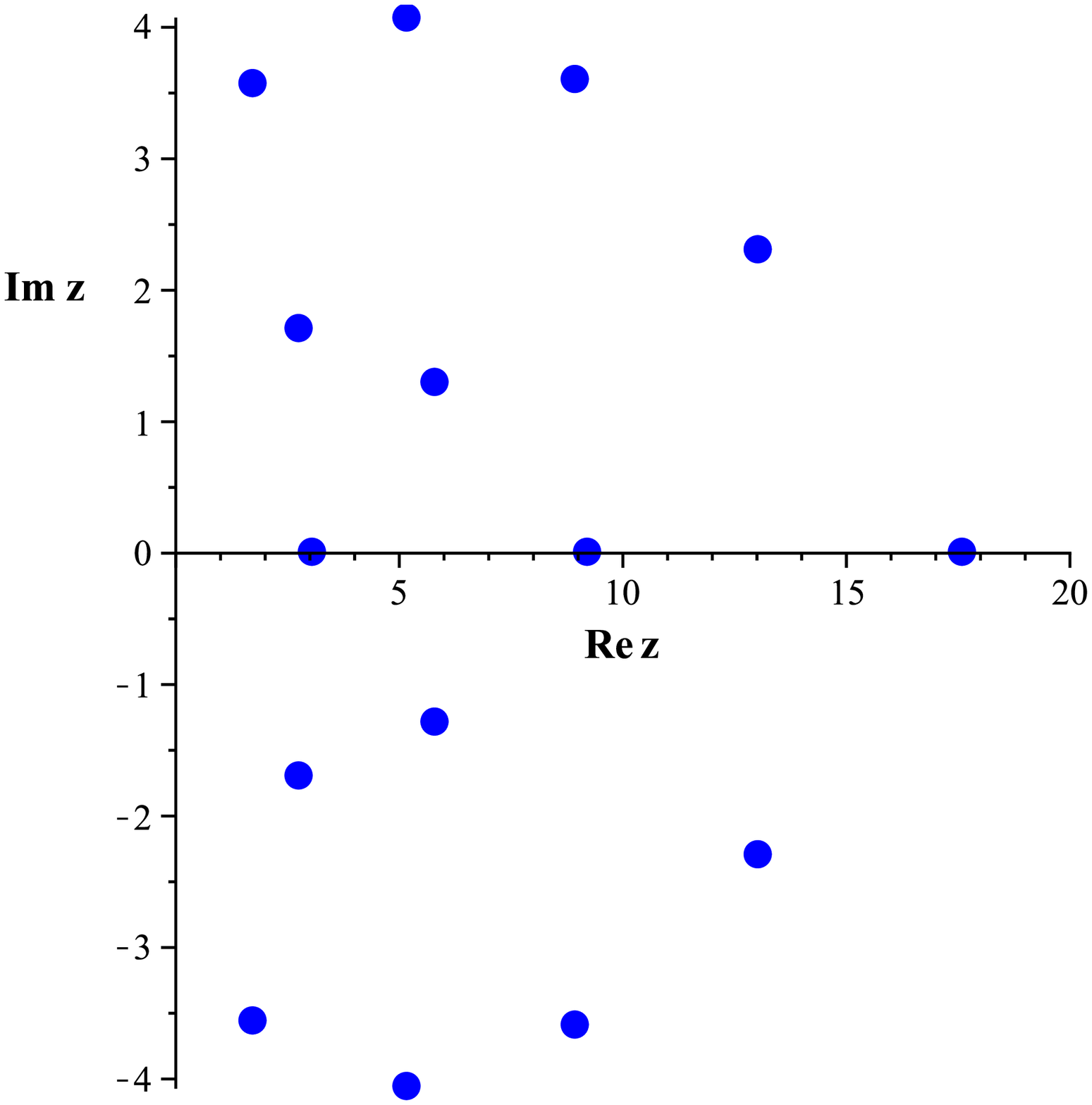,width=60mm}\label{}}
 \subfigure[$LR_{6}^{(2)}(z)$]{\epsfig{file=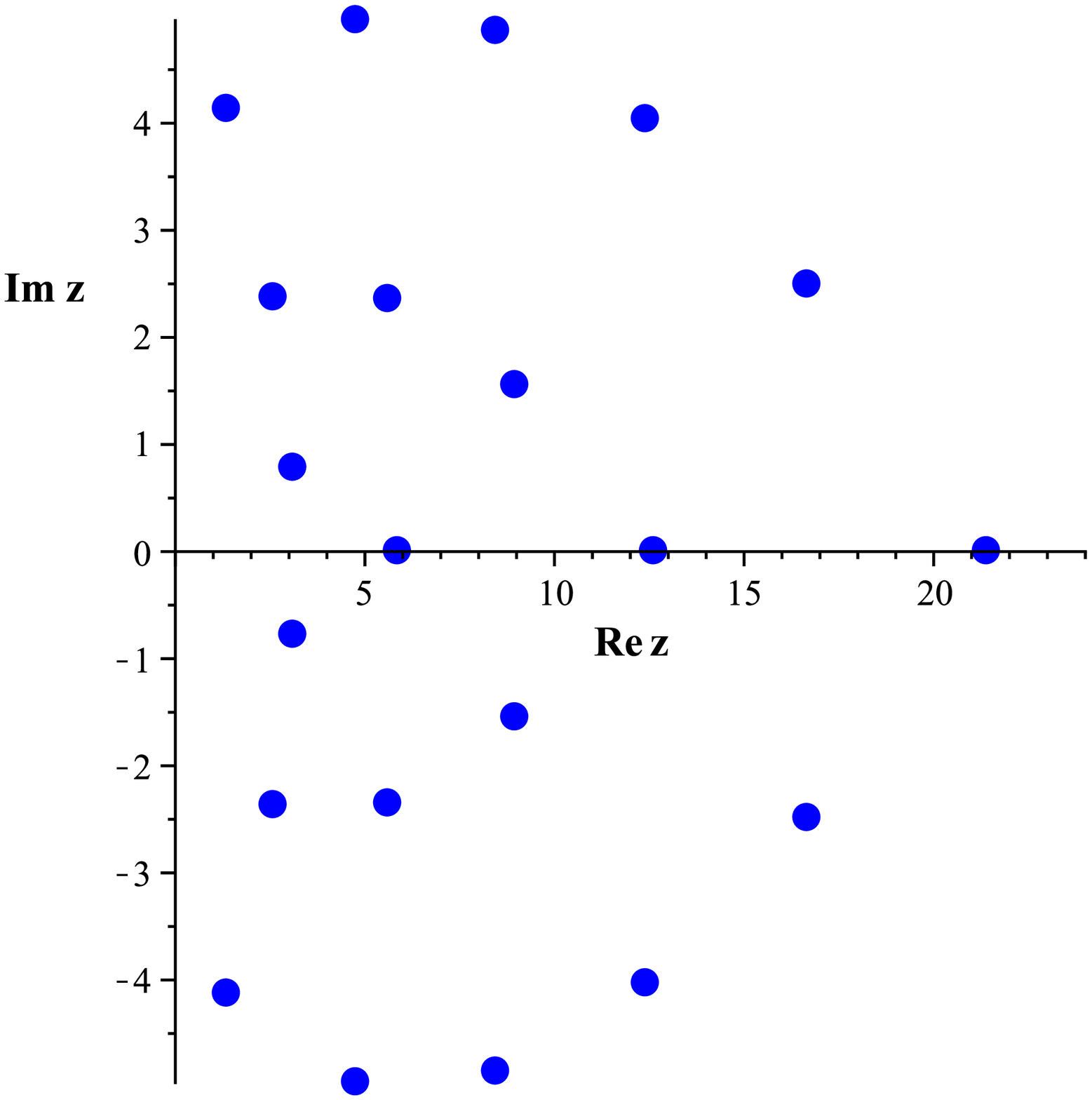,width=60mm}\label{}}}
 \centerline{
 \subfigure[$LR_{7}^{(2)}(z)$]{\epsfig{file=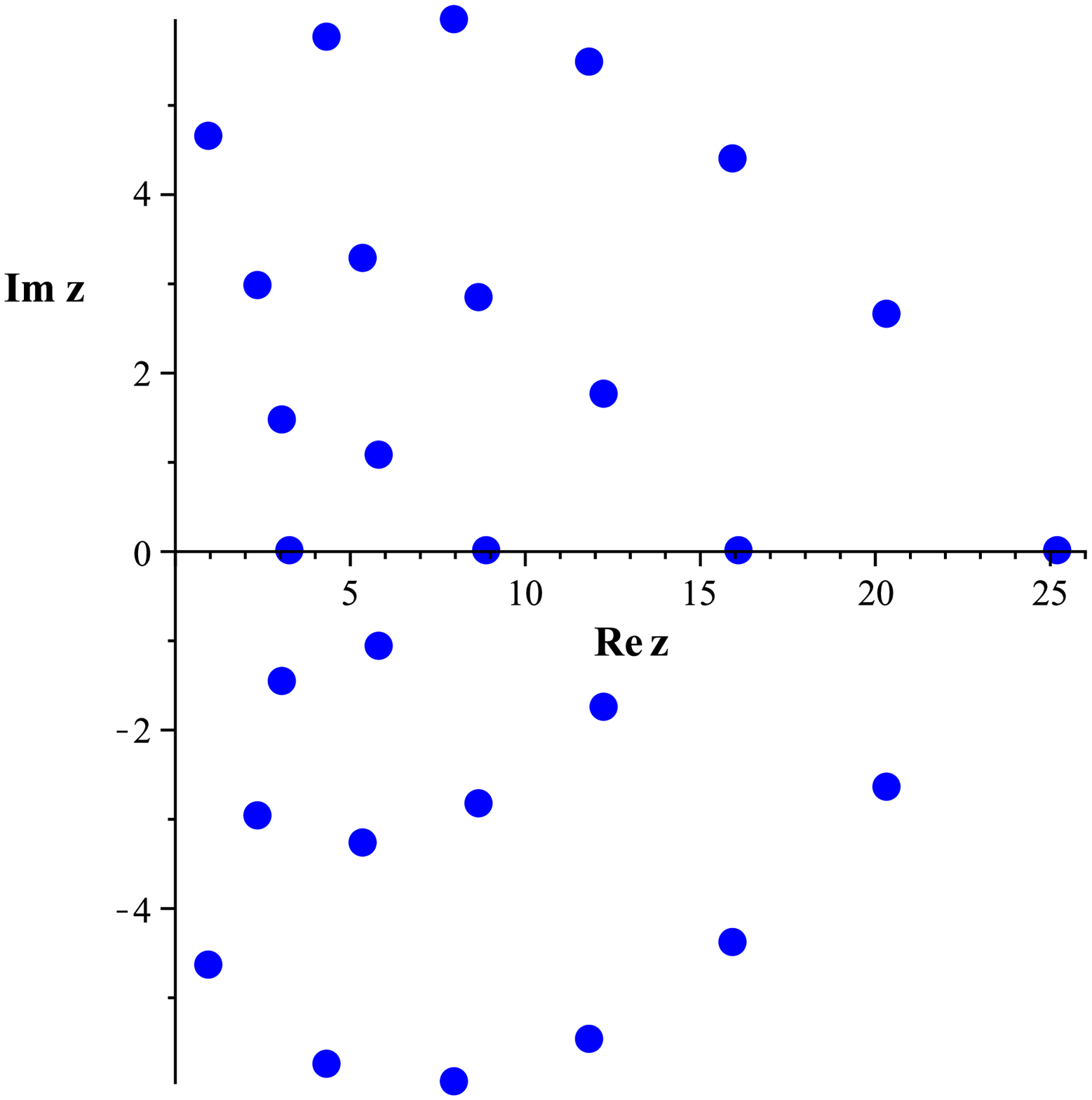,width=60mm}\label{}}
 \subfigure[$LR_{8}^{(2)}(z)$]{\epsfig{file=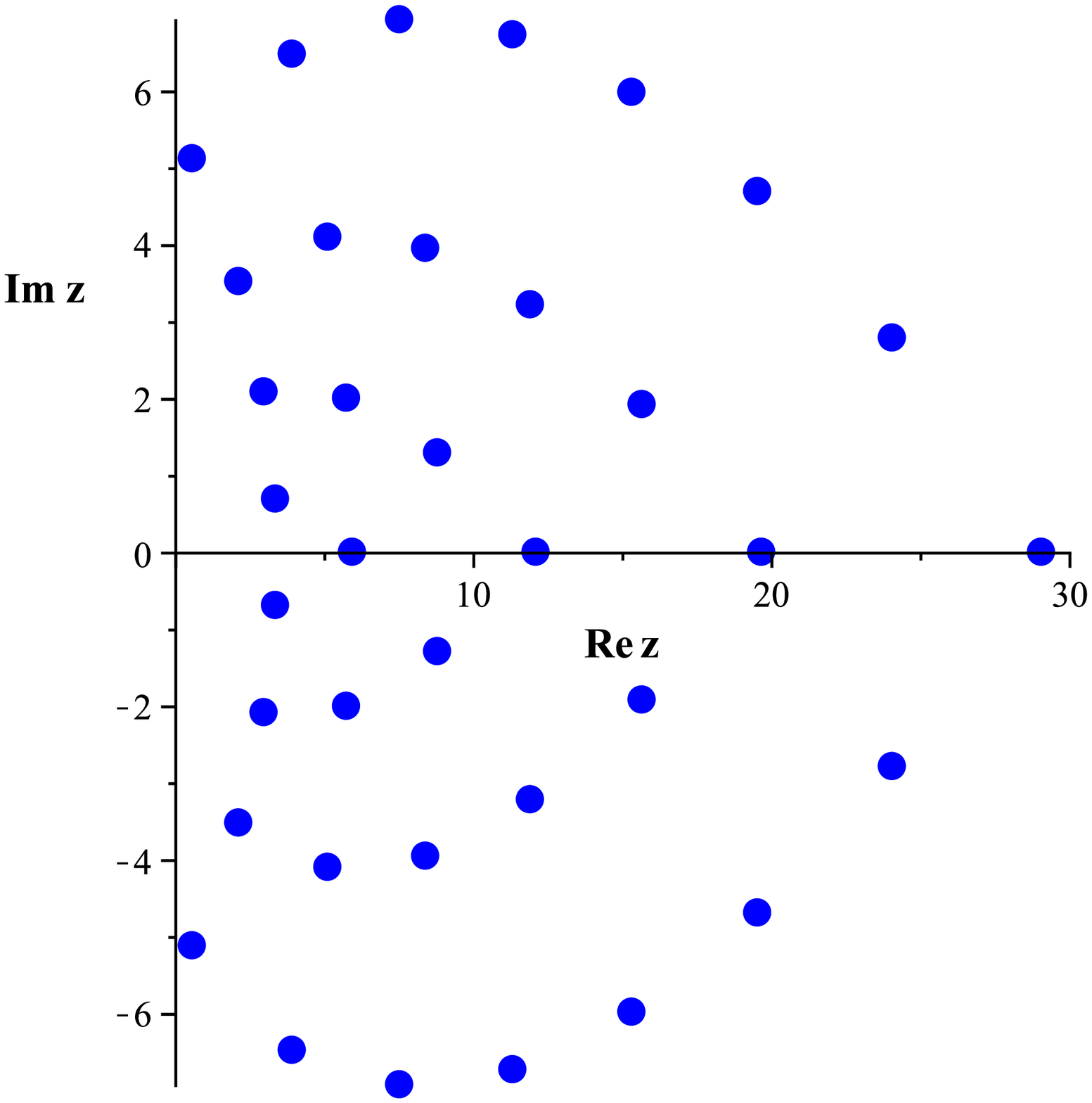,width=60mm}\label{}}}
   \caption{Roots of the polynomials $\{LR_{k}^{(\alpha)}(z)\}$.}
 \label{F:LR_1}
\end{figure}

First of all let us study the case when one of the polynomials ($\tilde{P}(z)$ or $\tilde{Q}(z)$) in \eqref{Equation_Tilde_PQ_BF} is a constant.   Setting $\tilde{Q}(z)=1$ in \eqref{Equation_Tilde_PQ_BF}, we obtain the equation for the polynomial $\tilde{P}(z)$
\begin{equation}
\begin{gathered}\label{Equation_Tilde_PQ_BF_Qeqv1}R(z)\tilde{P}_{zz}+\left[\frac{R_z(z)}{2}+U(z)\right]\tilde{P}_z-C\tilde{P}=0,
\end{gathered}
\end{equation}
where the constant $C$ is given by \eqref{Const_EQ_Background_Flow_Tilde} with $\deg \tilde{Q}=0$. The equation \eqref{Equation_Tilde_PQ_BF_Qeqv1} coincides with the equation for classical orthogonal polynomials (see \eqref{Ort_Pols_Equation}) provided that the polynomials $R(z)$, $U(z)$ are taken in the form \eqref{BF_exact} with $k=0$. It is well known that classical orthogonal polynomials do not have multiple roots and do not have common roots with the polynomial $R(z)$. Thus it follows from expression \eqref{New_Pols_Arbitrary_Circ} that the roots of any classical orthogonal polynomial give equilibrium positions of $M=\deg \tilde{P}$ point vortices with circulation $1$ in background flow \eqref{Background_Flow_Expression} with $R(z)$, $U(z)$ taken as \eqref{BF_exact} under the condition $k=0$. Analogously setting $\tilde{P}(z)=1$ in \eqref{Equation_Tilde_PQ_BF}, we obtain the equation for the polynomial $\tilde{Q}(z)$
\begin{equation}
\begin{gathered}\label{Equation_Tilde_PQ_BF_Peqv1}R(z)\tilde{Q}_{zz}+\left[\frac{R_z(z)}{2}-U(z)\right]\tilde{Q}_z-C\tilde{Q}=0,
\end{gathered}
\end{equation}
where the constant $C$ is given by \eqref{Const_EQ_Background_Flow_Tilde} with $\deg \tilde{P}=0$. This equation is exactly the equation for classical orthogonal polynomials (see \eqref{Ort_Pols_Equation}) if we take the polynomials $R(z)$, $U(z)$ in the form
\begin{equation}
\begin{gathered}
\label{BF_exact_alt}R(z)=\sigma(z),\quad  U(z)=\frac{\sigma_z(z)}{2}-\tau(z).
\end{gathered}
\end{equation}
Consequently from expression \eqref{New_Pols_Arbitrary_Circ} it follows that the roots of any classical orthogonal polynomial give equilibrium positions of $M=\deg \tilde{Q}$ point vortices with circulation $-1$ in background flow \eqref{Background_Flow_Expression} with the polynomials $U(z)$, $R(z)$ given by \eqref{BF_exact_alt}.

\begin{table}[t]%[h]
    \caption{Polynomials $\{LR_{k}^{(\alpha)}(z)\}$.} \label{t:LR_1}
  %  \center
       \begin{tabular}[pos]{|l|}
                \hline
                 $LR_0^{(\alpha)}(z)=1$\\
                $LR_1^{(\alpha)}(z)=z-\alpha-1$\\
        $LR_2^{(\alpha)}(z)=z^3-3(\alpha+2)z^2+3(\alpha+1)(\alpha+3)z-(\alpha+1)(\alpha+2)(\alpha+3)$\\
        $LR_3^{(\alpha)}(z)=z^6-6(\alpha+3)z^5+15(\alpha+2)(\alpha+4)z^4-10(\alpha+3)(2\alpha^2+12\alpha+13)z^3$\\\
        $\qquad \qquad +15(\alpha+1)(\alpha+3)^2(\alpha+5)z^2-6(\alpha+1)(\alpha+2)(\alpha+3)(\alpha+4)(\alpha+5)z$\\
        $\qquad \qquad +(\alpha+1)(\alpha+2)(\alpha+3)^2(\alpha+4)(\alpha+5)$\\
\hline
        \end{tabular}
\end{table}

\begin{figure}[t]
 \centerline{
 \subfigure[$UG_{6}(z)$]{\epsfig{file=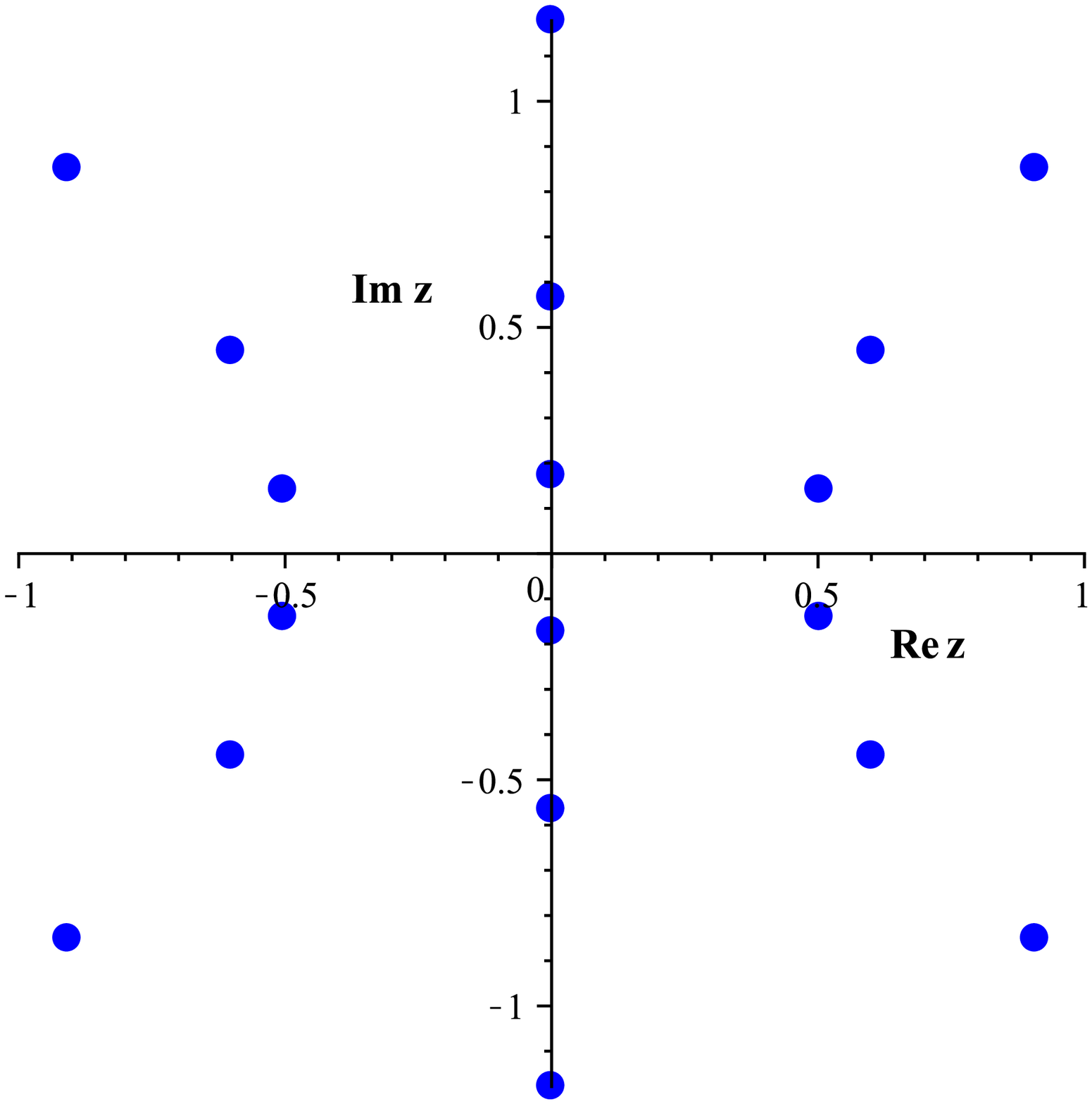,width=60mm}\label{}}
 \subfigure[$UG_{7}(z)$]{\epsfig{file=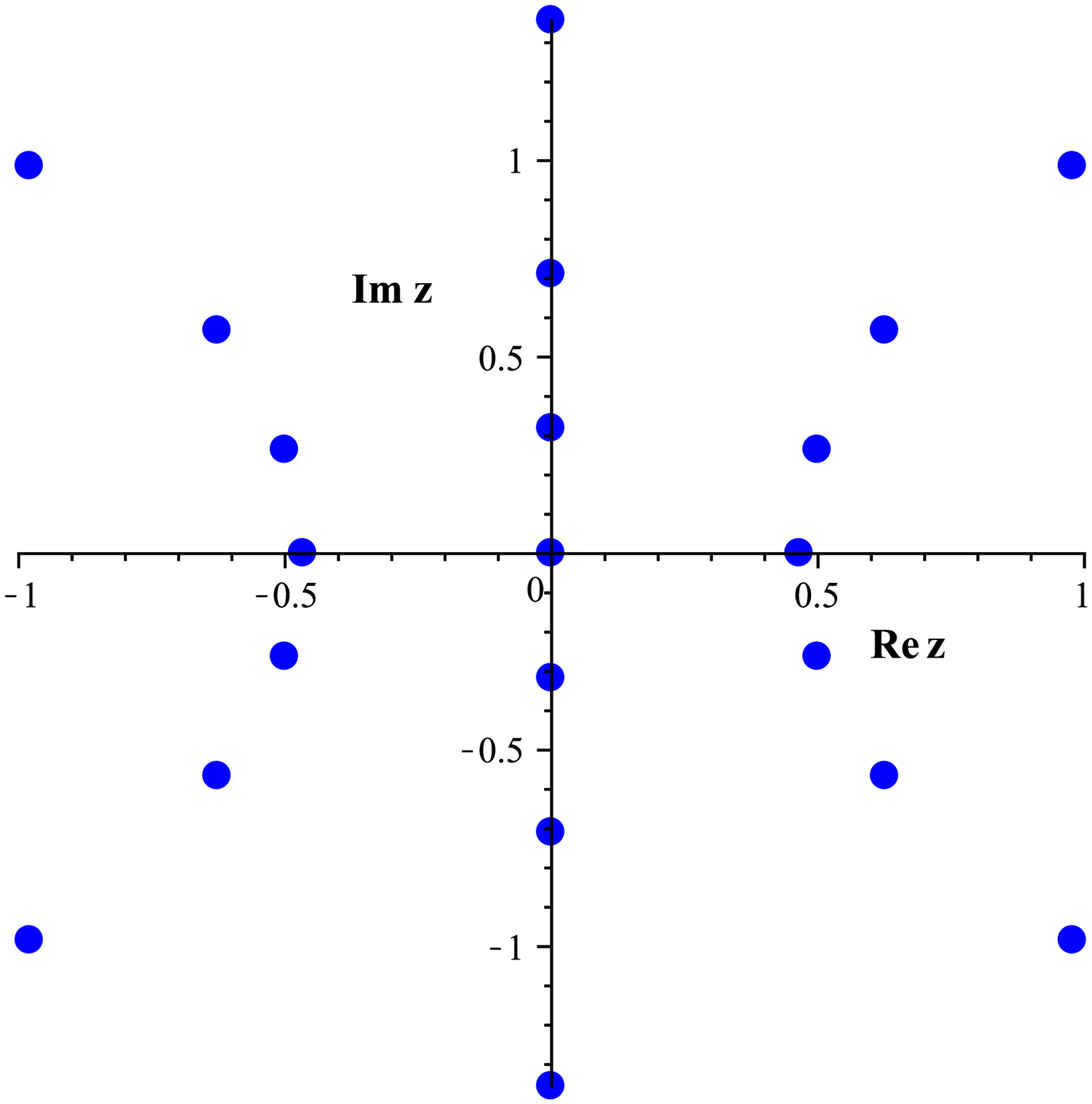,width=60mm}\label{}}}
 \centerline{
 \subfigure[$UG_{8}(z)$]{\epsfig{file=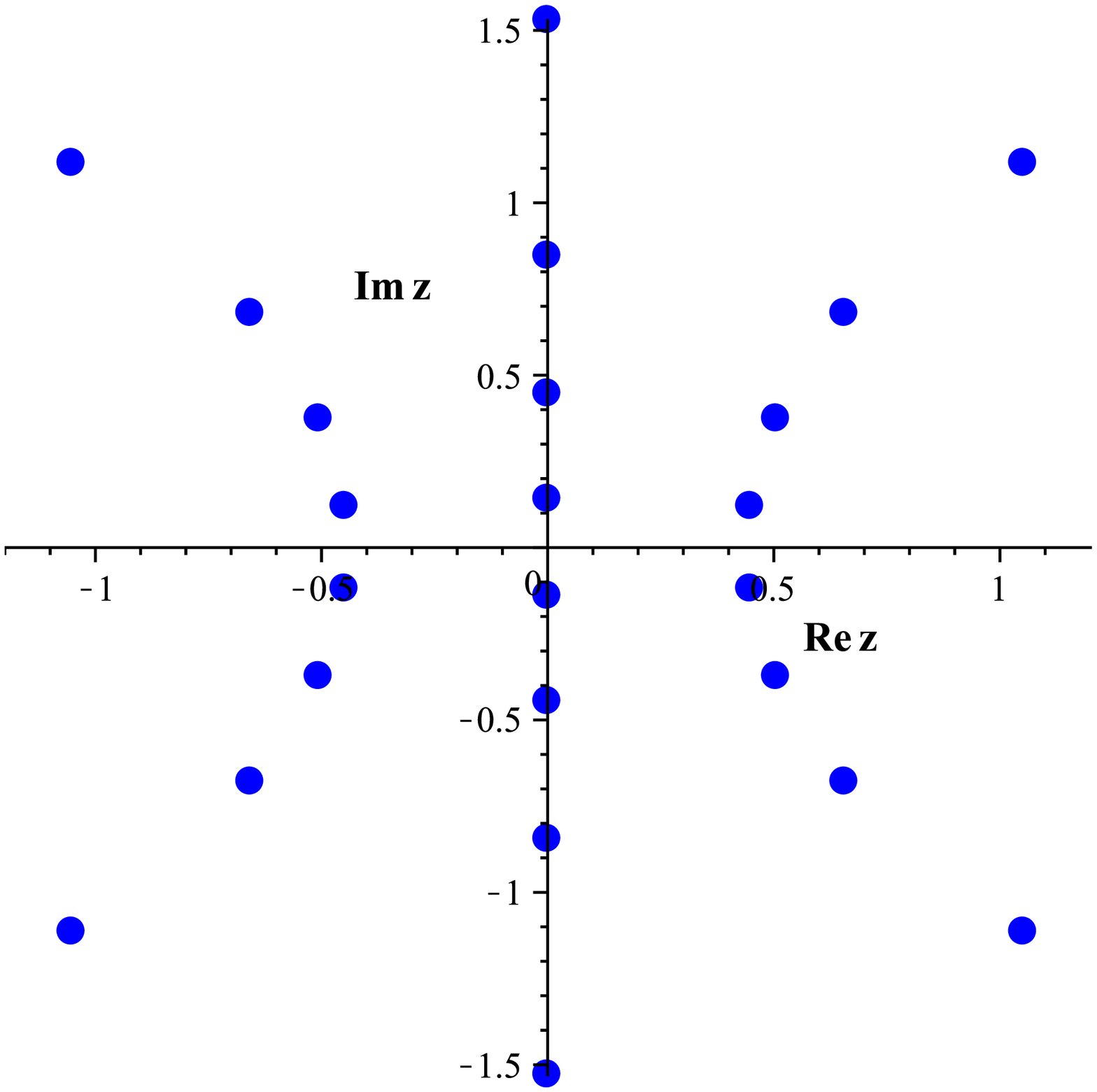,width=60mm}\label{}}
 \subfigure[$UG_{9}(z)$]{\epsfig{file=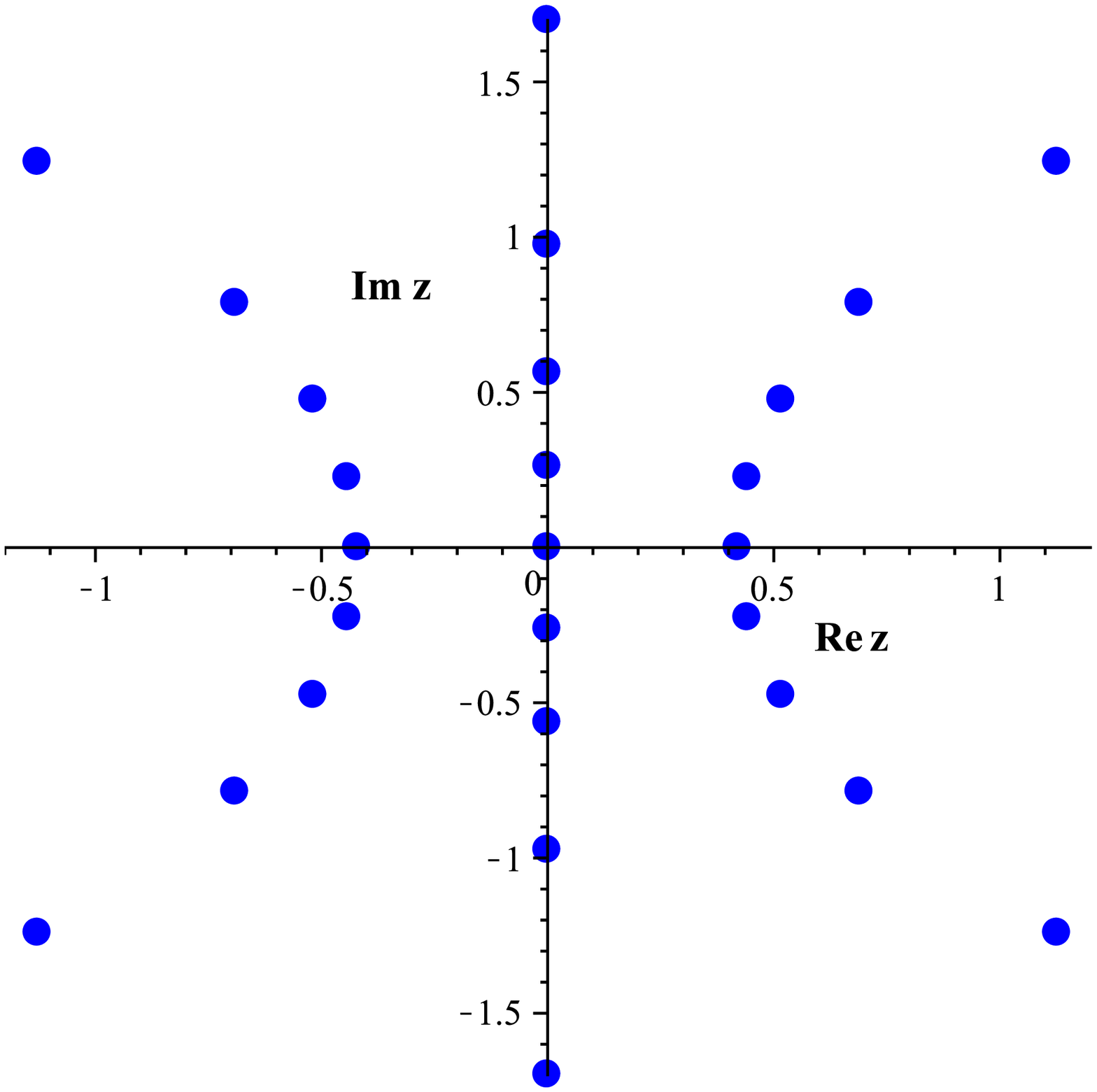,width=60mm}\label{}}}
   \caption{Roots of the polynomials $\{UG_{k}(z)\}$.}
 \label{F:UG_1}
\end{figure}

Further recalling the results of section \ref{Darboux_Orthogonal_Polynomials}, we consider polynomials that solve equation \eqref{Equation_Tilde_PQ_BF} and are Wronskians of classical orthogonal polynomials.  The sequence of Hermite polynomials satisfy the following second order differential equation
\begin{equation}
\begin{gathered}\label{Hermite_ODE}\psi_{zz}-2z\psi_z+2n\psi=0,\quad \psi =H_n(z).
\end{gathered}
\end{equation}
Thus, we have $\sigma(z)=1$, $\tau(z)=-2z$, $\lambda_n=2n$. The Hermite polynomials are orthogonal  with respect to the weight function  $\varrho(z)=\exp (-z^2)$ on the real line $(-\infty,\infty)$. Using expressions \eqref{Equation_Tilde_PQ_BF}, \eqref{BF_exact}, we see that the equation
\begin{equation}
\begin{gathered}
\label{Eqn_Determinants_Hermite_Polynomials}\tilde{P}_{zz}\tilde{Q}-2\tilde{P}_z\tilde{Q}_z+\tilde{P}\tilde{Q}_{zz}-2z\left\{\tilde{P}_z\tilde{Q}-\tilde{P}\tilde{Q}_z\right\}
+ 2\left(\deg \tilde{P} -\deg \tilde{Q}\right)\tilde{P}\tilde{Q}=0.
\end{gathered}
\end{equation}
possesses polynomial solutions  given by formula \eqref{Equation_Tilde_PQ_BF_ES} with $p_{i_k}=H_{i_k}(z)$. Originally this result appeared in \cite{Oblomkov01}. It is known  that polynomials that are Wronskians of the Hermite polynomials also arise in the theory of the Painlev\'{e} equations and their higher order analogues \cite{Clarkson01,Filipuk01}.

\begin{table}[t]%[h]
    \caption{Polynomials $\{UG_{k}(z)\}$.} \label{t:UG}
  %  \center
       \begin{tabular}[pos]{|l|}
                \hline
                \\
                 $UG_1(z)=z^3-\frac12 z$\\
                 \\
                 $UG_2(z)={z}^{6}-\frac34{z}^{4}+\frac{3}{16}{z}^{2}+\frac1{32}$\\
                 \\
                 $UG_3(z)={z}^{9}-\frac34{z}^{7}+\frac38{z}^{5}+\frac {5}{64}{z}^{3}-{\frac {15}{256}}z$\\
                 \\
                 $UG_4(z)={z}^{12}-\frac12{z}^{10}+{\frac {9}{16}}{z}^{8}+{\frac {9}{32}}{z}^{6}-{\frac {65}{256}}{z}^{4}+{\frac {15}{256}}{z}^{2}+{\frac {15}{4096}}$\\
                 \\
                 $UG_5(z)={z}^{15}+{\frac {15}{16}}{z}^{11}+{\frac {15}{16}}{z}^{9}-{\frac {
                 165}{256}}{z}^{7}+{\frac {147}{512}}{z}^{5}-{\frac {35}{4096}}{z}^{3}-{\frac {105}{8192}}z$\\
                 \\
                 $UG_6(z)={z}^{18}+\frac34{z}^{16}+{\frac {15}{8}}{z}^{14}+{\frac {175}{64}}{z}^{12}-{\frac {231}{256}}{z}^{10}+{\frac {231}{256}}{z}^{8}- {\frac {49}{2048}}{z}^{6}-$\\$\qquad\qquad-{\frac {735}{8192}}{z}^{4}+{\frac {735}{
                32768}}{z}^{2}+{\frac {49}{65536}}$\\
\\
\hline
        \end{tabular}
\end{table}

The sequence of Laguerre polynomials $L_n^{(\alpha)}(z)$ can be generated with the help of the following ordinary differential equation
\begin{equation}
\begin{gathered}\label{Laguerre_ODE}z\psi_{zz}+(\alpha+1-z)\psi_z+n\psi=0,\quad \psi= L_n^{(\alpha)}(z),\quad \alpha>-1.
\end{gathered}
\end{equation}
Note that sometimes these polynomials are called the generalized Laguerre polynomials whenever $\alpha\neq 0$, $\alpha>-1$. In this case we get $\sigma(z)=z$, $\tau(z)=\alpha+1-z$, $\lambda_n=n$. The Laguerre polynomials are orthogonal with respect to the weight function $\varrho(z)=z^{\alpha}\exp (-z)$ on the real interval $[0,\infty)$. By means of expressions \eqref{Equation_Tilde_PQ_BF}, \eqref{BF_exact} we obtain that the equation
\begin{equation}
\begin{gathered}
\label{Eqn_Determinants_Laguerre_Polynomials}z\left\{\tilde{P}_{zz}\tilde{Q}-2\tilde{P}_z\tilde{Q}_z+\tilde{P}\tilde{Q}_{zz}\right\}+\left(\alpha+k+\frac12 -z\right)\left\{\tilde{P}_z\tilde{Q}-\tilde{P}\tilde{Q}_z\right\}\\
+ \frac{1}{2}\left\{\tilde{P}_z\tilde{Q}+\tilde{P}\tilde{Q}_z\right\}+\left(\deg \tilde{P} -\deg \tilde{Q}\right)\tilde{P}\tilde{Q}=0.
\end{gathered}
\end{equation}
has polynomial solutions  given by formula \eqref{Equation_Tilde_PQ_BF_ES} with $p_{i_k}=L_{i_k}^{(\alpha)}(z)$. Let us consider an example. Two neighbor polynomials from the sequence
\begin{equation}
\begin{gathered}
\label{Laguerre_Polynomials_Example}LR_{k}^{(\alpha)}(z)=\delta_k W[L_1^{(\alpha)},L_3^{(\alpha)},\ldots , L_{2k-1}^{(\alpha)}],\quad k\in \mathbb{N} \\
LR_{0}^{(\alpha)}(z)=1,\hfill
\end{gathered}
\end{equation}
i.e. $\tilde{P}(z)=LR_{k+1}^{(\alpha)}(z)$, $\tilde{Q}(z)=LR_{k}^{(\alpha)}(z)$ solve equation \eqref{Eqn_Determinants_Laguerre_Polynomials}. This statement for the pair $\tilde{P}(z)=LR_{1}^{(\alpha)}(z)$, $\tilde{Q}(z)=LR_{0}^{(\alpha)}(z)$ follows from equation \eqref{Equation_Tilde_PQ_BF_Qeqv1} and remarks after it.
Without loss of generality, we choose the constant $\delta_k$ in such a way that the resulting polynomial is monic. First few polynomials from the sequence \eqref{Laguerre_Polynomials_Example} are given in table \ref{t:LR_1}.  Using relation \eqref{Degreees_W_Orthogonal_Polynomials}, we find the degree of the polynomial $LR_{k}^{(\alpha)}(z)$
\begin{equation}
\begin{gathered}
\label{Laguerre_Polynomials_Example_Degree}\deg LR_{k}^{(\alpha)}(z)=\frac{k(k+1)}{2}.
\end{gathered}
\end{equation}
Interestingly, that roots of the polynomials $LR_{k}^{(\alpha)}(z)$ form highly regular structures in the complex plane. Examples of plots are given in figure \ref{F:LR_1}.

\begin{figure}[t]
 \centerline{
 \subfigure[$LG_{3}(z)$]{\epsfig{file=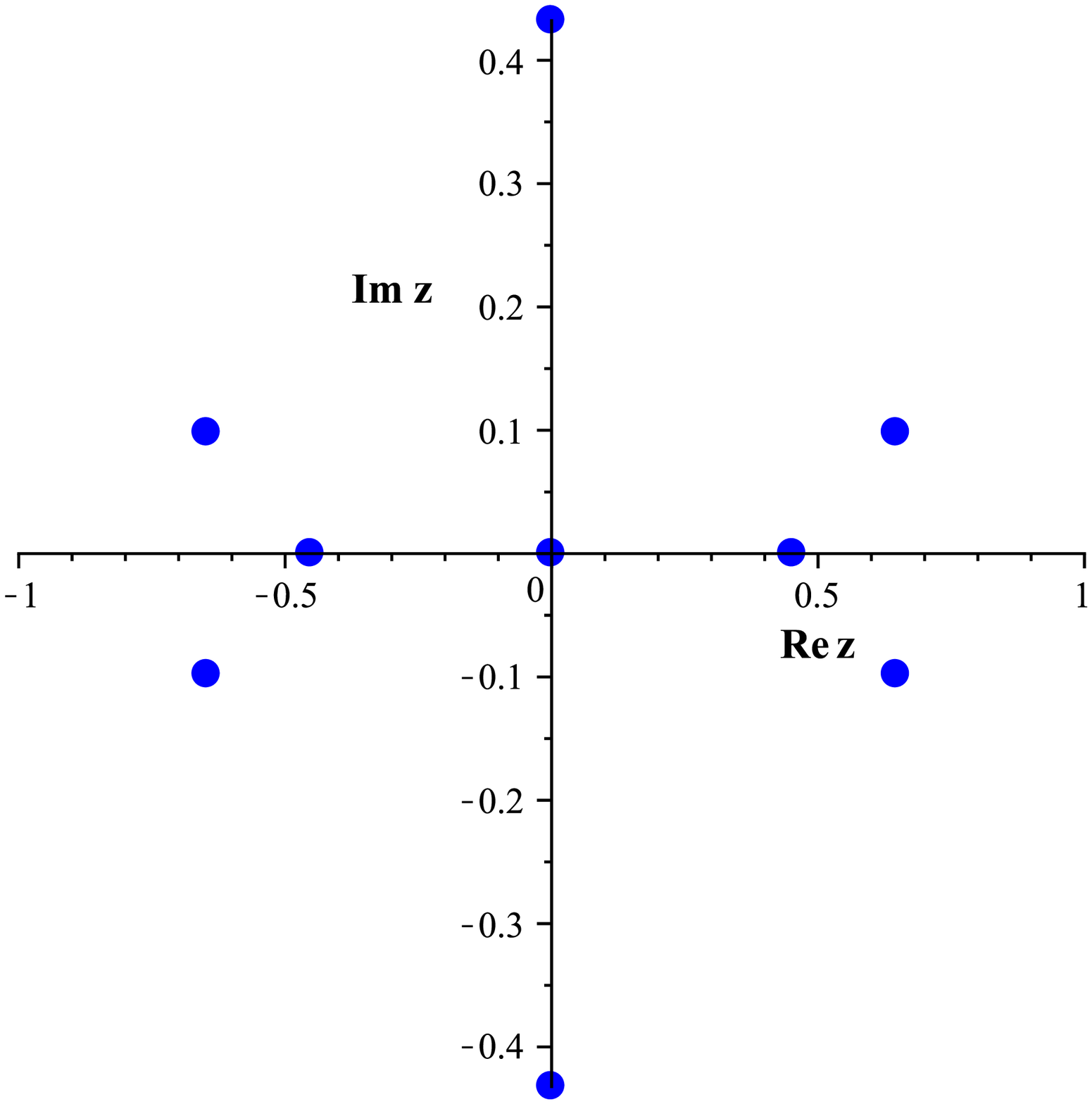,width=60mm}\label{}}
 \subfigure[$LG_{4}(z)$]{\epsfig{file=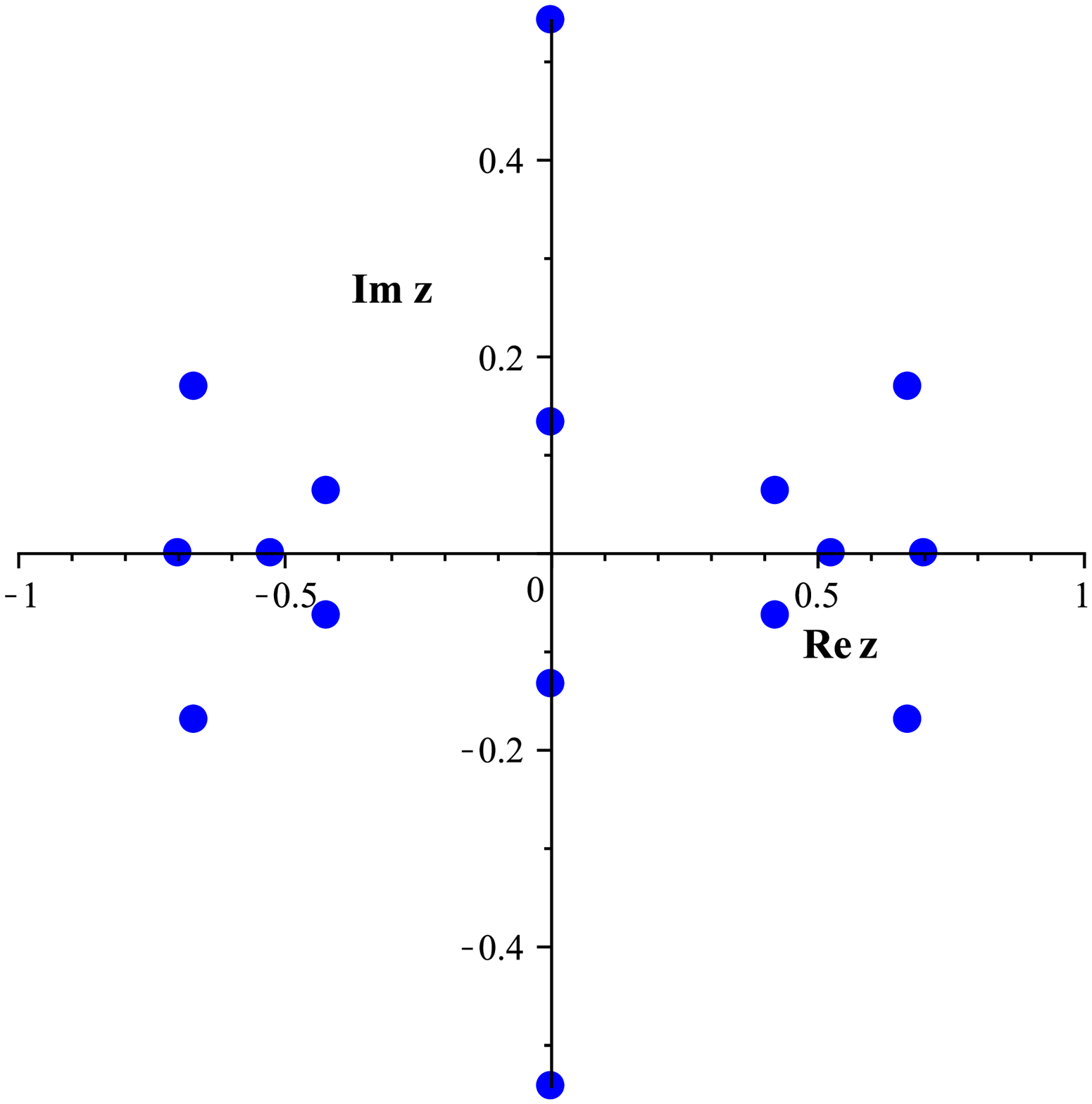,width=60mm}\label{}}}
 \centerline{
 \subfigure[$LG_{5}(z)$]{\epsfig{file=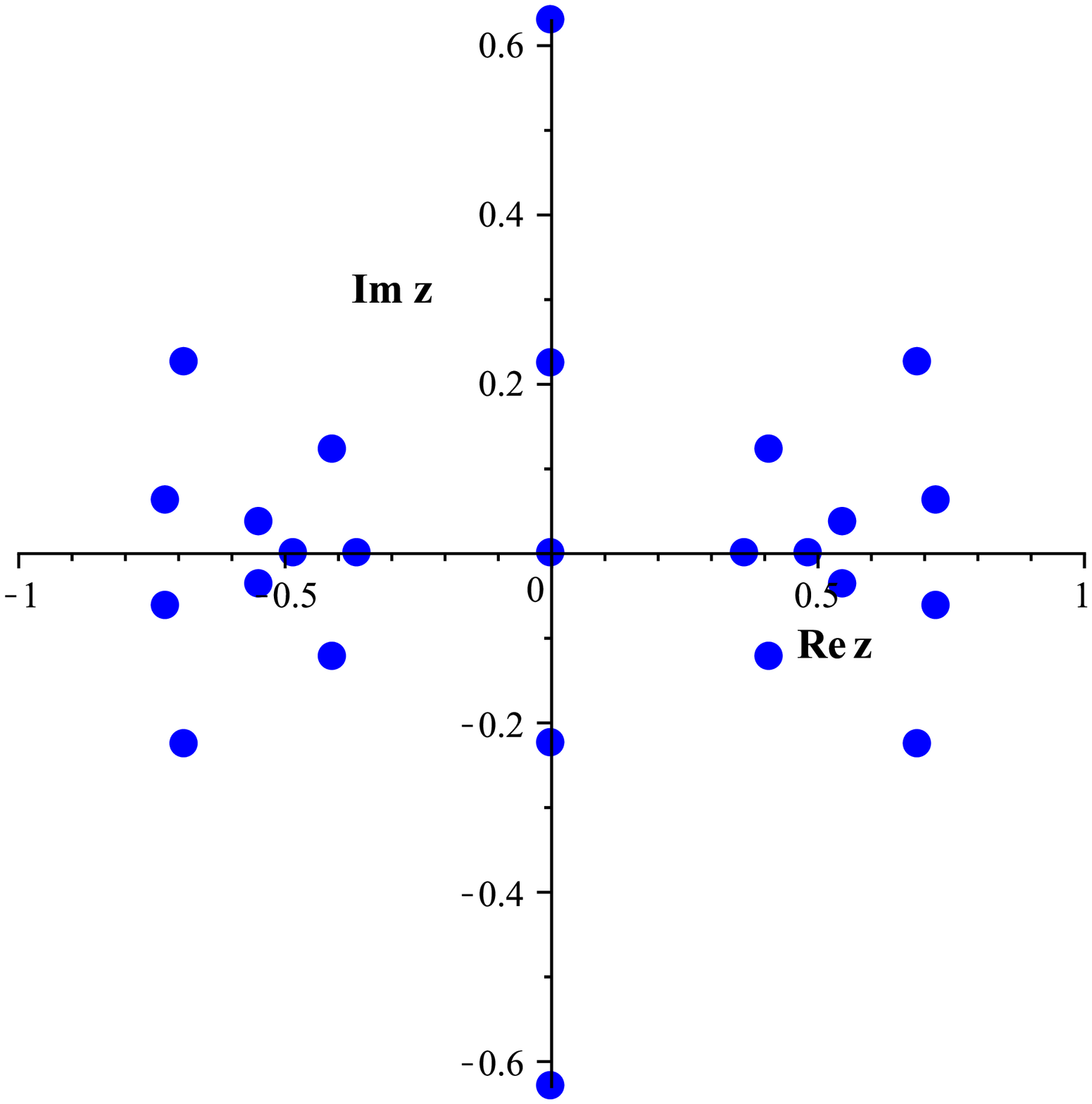,width=60mm}\label{}}
 \subfigure[$LG_{6}(z)$]{\epsfig{file=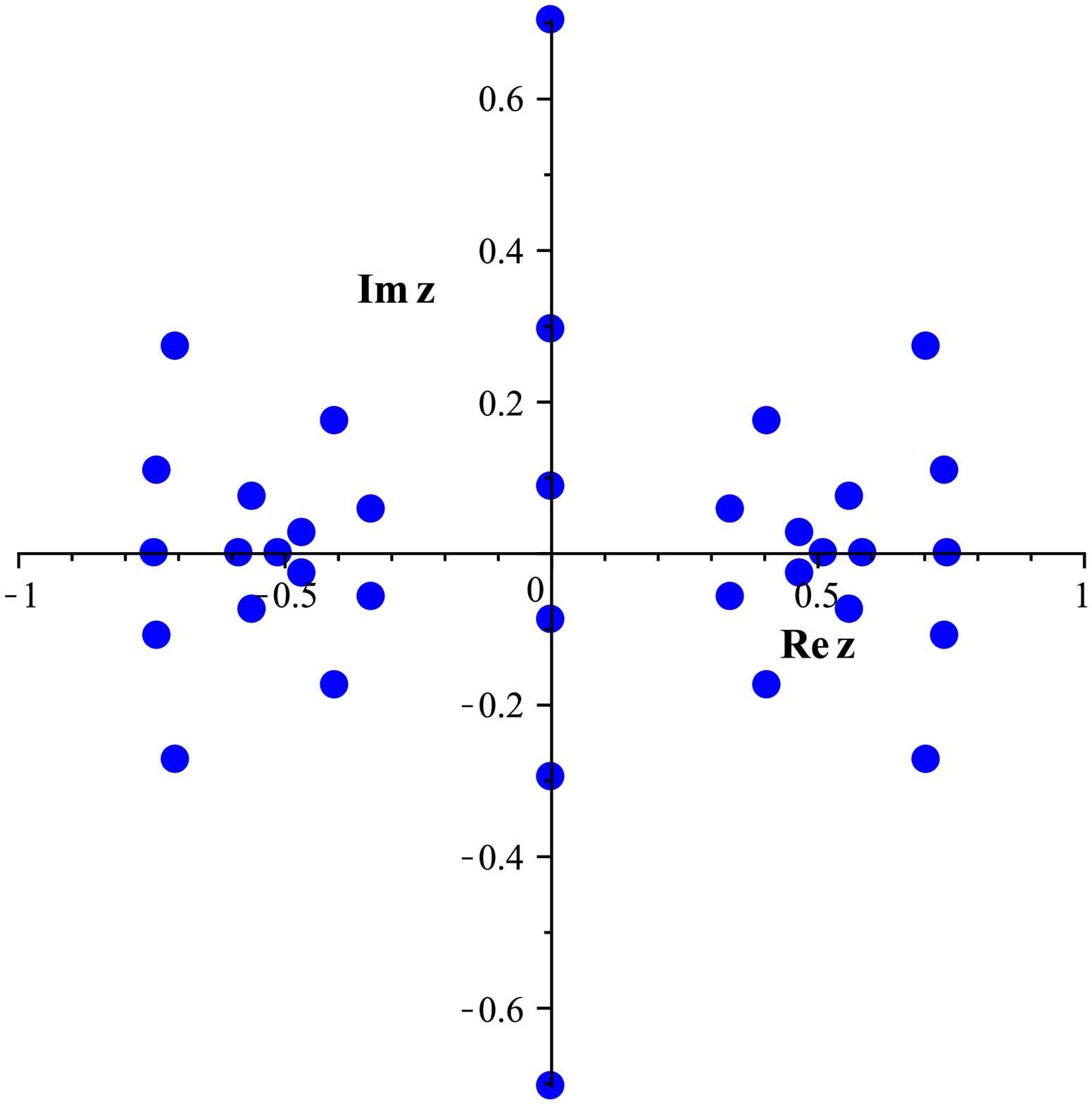,width=60mm}\label{}}}
   \caption{Roots of the polynomials $\{LG_{k}(z)\}$.}
 \label{F:LG_1}
\end{figure}

Classical polynomials orthogonal on a finite interval, which is usually taken as the real interval $[-1,1]$, are the Jacobi polynomials $P_n^{(\alpha,\beta)}(z)$ and their partial cases including the Chebyshev polynomials of the first kind $T_n(z)$ and of the second kind $U_n(z)$, the Gegenbauer polynomials $C_n^{(\lambda)}(z)$, and the Legendre polynomials $P_n(z)$. The Jacobi polynomials satisfy the following second order equation
\begin{equation}
\begin{gathered}\label{Jacobi_ODE}(1-z^2)\psi_{zz}+[\beta-\alpha-(\alpha+\beta+2)z]\,\psi_z+n(n+\alpha+\beta+1)\psi=0,\\
 \psi= P_n^{(\alpha,\beta)}(z),\quad \alpha>-1,\quad \beta>-1.
\end{gathered}
\end{equation}
Consequently, we have $\sigma(z)=1-z^2$, $\tau(z)=\beta-\alpha-(\alpha+\beta+2)z$, $\lambda_n=n(n+\alpha+\beta+1)$. Note that the Jacobi  polynomials are orthogonal with respect to the weight function $\varrho(z)=(1-z)^{\alpha}(1+z)^{\beta}$ on the real interval $[-1,1]$. Using expressions \eqref{Equation_Tilde_PQ_BF}, \eqref{BF_exact}, we find the differential equation
\begin{equation}
\begin{gathered}
\label{Eqn_Determinants_Jacobi_Polynomials}(1-z^2)\left\{\tilde{P}_{zz}\tilde{Q}-
2\tilde{P}_z\tilde{Q}_z+\tilde{P}\tilde{Q}_{zz}\right\}+\\
+\left[\beta-\alpha-(\alpha+\beta+2k+1)z\right] \left\{\tilde{P}_z\tilde{Q} -\tilde{P}\tilde{Q}_z\right\}-z\left\{\tilde{P}_z\tilde{Q}+\tilde{P}\tilde{Q}_z\right\}+\\+
\left(\deg \tilde{P} -\deg \tilde{Q}\right)\left(\deg \tilde{P} -\deg \tilde{Q}+\alpha+\beta+2k+1\right)\tilde{P}\tilde{Q}=0,
\end{gathered}
\end{equation}
which possesses polynomial solutions in the form \eqref{Equation_Tilde_PQ_BF_ES} with $p_{i_k}=P_{i_k}^{(\alpha,\beta)}(z)$. In the case $\alpha=\beta=1/2$ the Jacobi  polynomials $P_n^{(\alpha,\beta)}(z)$ become the Chebyshev polynomials of the second kind $U_n(z)$.  As an example we consider the following sequence of polynomials
\begin{equation}
\begin{gathered}
\label{Chebyshev_Polynomials_Example}UG_{k}(z)=\gamma_k W[U_3,U_4,\ldots,U_{k+2}],\quad k\in \mathbb{N}.
\end{gathered}
\end{equation}
Two neighbor polynomials from this sequence, in other words the polynomials $\tilde{P}(z)=UG_{k+1}(z)$, $\tilde{Q}(z)=UG_{k}(z)$, satisfy equation \eqref{Eqn_Determinants_Jacobi_Polynomials} with $\alpha=\beta=1/2$. In expression \eqref{Chebyshev_Polynomials_Example} we choose the constant $\gamma_k$ in such a way that the corresponding polynomial is monic. With the help of relation \eqref{Degreees_W_Orthogonal_Polynomials} we calculate the degree of the polynomial $UG_{k}(z)$
\begin{equation}
\begin{gathered}
\label{Chebyshev_Polynomials_Example_Degree}\deg UG_{k}(z)=3k.
\end{gathered}
\end{equation}
First few polynomials $UG_{k}(z)$ in explicit form are given in table \ref{t:UG}. Again the roots of the polynomials $UG_k(z)$ form highly regular structures in the complex plane. Several examples are plotted in figure \ref{F:UG_1}.

\begin{table}[t]%[h]
    \caption{Polynomials $\{LG_{k}(z)\}$.} \label{t:LG}
  %  \center
       \begin{tabular}[pos]{|l|}
                \hline
                \\
                 $LG_1(z)=z$\\
                 \\
                 $LG_2(z)={z}^{4}-\frac27{z}^{2}-\frac{1}{35}$\\
                 \\
                 $LG_3(z)={z}^{9}-{\frac {76}{91}}{z}^{7}+{\frac {114}{715}}{z}^{5}+{\frac {4}{143}}{z}^{3}-{\frac {1}{143}}z$\\
                 \\
                 $LG_4(z)={z}^{16}-{\frac {2840}{1729}}{z}^{14}+{\frac {45964}{46189}}{z}^{12}-{\frac {856}{4199}}{z}^{10}-{\frac {1970}{46189}}{z}^{8}+{\frac {16408}{600457}}{z}^{6}-$\\$\qquad\qquad-{\frac {140}{31603}}{z}^{4}+{\frac {1400}{6605027}}{z}^{2}+{\frac {35}{6605027}}$\\
                 \\
\hline
        \end{tabular}
\end{table}

The Legendre polynomials $P_n(z)$ are special cases of the Jacobi  polynomials and correspond to zero values of the parameters $\alpha$, $\beta$. Let us consider the following sequence of polynomials
\begin{equation}
\begin{gathered}
\label{Legendre_Polynomials_Example}LG_{k}(z)=\omega_k W[P_1,P_4,\ldots,P_{3k-2}],\quad k\in \mathbb{N}.
\end{gathered}
\end{equation}
Two neighbor polynomials from this sequence $\tilde{P}(z)=LG_{k+1}(z)$, $\tilde{Q}(z)=LG_{k}(z)$, satisfy equation \eqref{Eqn_Determinants_Jacobi_Polynomials} with $\alpha=\beta=0$. In expression \eqref{Legendre_Polynomials_Example} we take the constant $\omega_k$ in such a way that the corresponding polynomial is monic. By means of relation \eqref{Degreees_W_Orthogonal_Polynomials} we derive the degree of the polynomial $LG_{k}(z)$
\begin{equation}
\begin{gathered}
\label{Legendre_Polynomials_Example_Degree}\deg LG_{k}(z)=k^2.
\end{gathered}
\end{equation}
Several polynomials $LG_{k}(z)$ in explicit form are presented in table \ref{t:LG}. The roots of the polynomials $LG_k(z)$ form highly regular structures in the complex plane. Several examples are given in figure \ref{F:LG_1}.

Finally, we would like to mention that our investigation establishes an additional connection between the point vortex theory and the theory of classical orthogonal polynomials. We hope that our results will be useful in further studying of point vortex equilibria.

\section{Conclusion}

In this article we have studied the problem of finding stationary configurations of point vortices with generic choice of circulations in a background flow. We have found differential equations satisfied by generating polynomials of vortex arrangements and have shown that these equations can be reduced to a single one. We have studied the latter equation and have derived polynomial solutions expressed as Wronskians of classical orthogonal polynomials. This result was obtained with the help of an approach based on the technique of Darboux transformation. In details we considered several examples involving Hermite, Laguerre, Chebyshev, and Legendre polynomials.

\section{Acknowledgements}

This research was partially supported by Federal Target Programm
"Research and Scientific--Pedagogical Personnel of Innovation
in Russian Federation on 2009-–2013".

\end{document}